\documentclass{ws-ijmpcs}

\usepackage{graphicx}
\usepackage{amssymb}
\usepackage{amsmath}
\usepackage{bm}
\usepackage{slashed}
\usepackage{datetime}
\usepackage{mymacros}
\usepackage{psfrag}

\def\trimmarks{NIKHEF 2013-031}

\begin{document}

\markboth{A. Signori, A. Bacchetta, M. Radici}
{Flavor dependence of unpolarized TMDs from semi-inclusive pion production}

%
\catchline{}{}{}{}{}
%

\title{FLAVOR DEPENDENCE OF UNPOLARIZED TMDS \\ FROM SEMI-INCLUSIVE PION PRODUCTION}

\author{ANDREA SIGNORI\footnote{Speaker at the QCD Evolution Workshop 2013.}}
\address{Nikhef Theory Group and Department of Physics and Astronomy, VU University Amsterdam\\ De Boelelaan 1081, NL-1081 HV Amsterdam, the Netherlands\\
asignori@nikhef.nl}

\author{ALESSANDRO BACCHETTA}
\address{INFN Sezione di Pavia and Dipartimento di Fisica, Universit\`a di Pavia \\ via Bassi 6, 27100 Pavia, Italy\\
alessandro.bacchetta@unipv.it}

\author{MARCO RADICI}
\address{INFN Sezione di Pavia \\
via Bassi 6, 27100 Pavia, Italy\\
marco.radici@pv.infn.it}

\maketitle


\begin{abstract}
Recent data from semi-inclusive deep inelastic scattering collected by the HERMES collaboration allow for the first time to discuss how the transverse-momentum dependence of unpolarized distribution and fragmentation functions is affected by the flavor of the involved partons. 
A model built with flavor-dependent Gaussian transverse-momentum distributions fits data better than the same flavor-independent model.
The current analysis is performed for totally unpolarized scattering of leptons off protons and deuterons, with detected pions in the final state.
There are convincing indications of flavor dependence in the fragmentation functions, while for parton distribution functions the evidence is weaker.
\end{abstract}

\ccode{PACS numbers: 13.60.-r, 13.87.Fh, 14.20.Dh, 14.65.Bt}

\section{Introduction}	
In the last decade transverse-momentum-dependent distribution and fragmentation functions (TMDs) have gained increasing attention both theoretically and experimentally, the latter especially because of the large amount of emerging data from semi-inclusive deep-inelastic scattering\cite{Airapetian:2012ki,Adolph:2013stb} (SIDIS).


In this work the transverse-momentum dependence of TMDs is chosen to be Gaussian.
This hypothesis simplifies the analysis and is adequate to these first studies, but a more accurate functional form may eventually be needed.


The original feature of our analysis is that we consider the possibility that the width of the TMDs is different for different quark flavors. This approach leads to a distinction mainly among valence and sea quarks and between favored and unfavored fragmentation processes (see Sec. \ref{s:results} and Sec. \ref{s:conclusions}).

\section{Unpolarized SIDIS cross section and multiplicity}
For a review of the theoretical framework underlying this analysis we refer to Ref. \refcite{Signori:2013mda}.\\
In one-hadron inclusive DIS a lepton scatters off a nucleon and one hadron is identified in the final state:
\begin{equation}
\ell + H \to \ell' + h + X \ .
\label{sidis}
\end{equation}
$\ell$, $\ell'$ denote the incoming and outgoing lepton respectively, $H$ the nucleon target, and $h$ the detected hadron.
Let's define the transverse momenta involved in our analysis:\\

\noindent
\begin{tabular}{ll}
Momentum  & Physical description \\
\hline
$\bm{k}_\perp$ & intrinsic partonic transverse momentum  
\\ 
 $\bm{P}_\perp$ & transverse momentum of final hadron w.r.t. fragmenting parton
\\
$\bm{P}_{hT}$ & transverse momentum of final hadron w.r.t. virtual photon
\end{tabular} \\

We will consider unpolarized scattering integrated over the azimuthal angle $\phi_h$ of the detected hadron, in one-photon exchange approximation, with no QCD corrections, in the limit $\bm{P}_{hT}^2 \ll Q^2$ (small transverse momenta) and $M^2 \ll Q^2$ (leading twist), being $Q^2$ the hard scale of the process and $M$ the mass of the target hadron. 
\subsection{Hadron multiplicities}
The available data sets deal with hadron multiplicities in SIDIS, namely the differential number of hadrons produced per DIS event. Casting this concept in terms of cross sections we have:
\begin{equation}
m_H^h (x,z,Q^2,\bm{P}_{hT}^2) = \frac{d^{(4)} \sigma_H^h / dx dQ^2 dz d\bm{P}_{hT}^2 }{ d^{(2)}\sigma_{\text{DIS}} / dx dQ^2 }\ .
\label{e:multiplicity}
\end{equation}
$d^{(4)}\sigma_H^h$ is the differential cross section for SIDIS (the superscript $h$ is related to the detected hadron, the subscript $H$ to the target involved in the scattering; $x$, $z$ are the usual light-cone momentum fractions), while $d^{(2)}\sigma_{\text{DIS}}$ is the corresponding inclusive one. 
The leading-twist and leading-order unpolarized SIDIS cross section is built in terms of the unpolarized transverse structure function $F_{UU,T}$\cite{Bacchetta:2006tn}:
\begin{equation}
\frac{d^{(4)}\sigma}{dx\ dQ^2 \ dz \ d\bm{P}_{hT}^2} = \frac{\pi \alpha^2}{x Q^4} \biggl[ 1+ \biggl( 1-\frac{Q^2}{x(s-M^2)} \biggr) \biggr] \ F_{UU,T}(x,z,Q^2,\bm{P}_{hT}^2)\ ,
\label{e:cross_LT_LO}
\end{equation} 
where $\alpha$ is the fine structure constant.
For $F_{UU,T}$ we rely on the factorized formula for low transverse momentum SIDIS:
%
\begin{equation} 
\label{F_UUT_simpl}
F_{UU ,T} (x,z,Q^2,\bm{P}_{hT}^2) =  \sum_a e_a^2\ \bigg[f_1^a(x,\bm{k}_{\perp}^2,Q^2) \otimes D_1^{a \to h}(z,\bm{P}_{\perp}^2, Q^2) \bigg] \ ,
\end{equation}
where $a$ is the flavor index\footnote{In summing over the flavor index up and down contributions will consist of two parts: one for valence quarks and one for sea quarks.} and the convolution is defined in Ref. \refcite{Bacchetta:2006tn}. For a precise definition of the involved variables we refer to Ref. \refcite{Signori:2013mda}. 

\subsection{Flavor-dependent Gaussian TMDs}
The flavor-dependent Gaussian hypothesis consists in assuming flavor-dependent Gaussian isotropic behavior for the transverse momentum in the distribution function $f_1^a$ and the fragmentation function $D_1^{a \to h}$:

\begin{align}
f_1^a(x,Q^2,\bm{k}_\perp^2) &= \frac{f_1^a(x,Q^2)}{\pi \langle \bm{k}_{\perp,a}^2 \rangle}e^{-\bm{k}_\perp^2/\langle \bm{k}_{\perp,a}^2 \rangle} ,
\nonumber
\\
D_1^{a \to h}(z,Q^2,\bm{P}_\perp^2) &= \frac{D_1^{a \to h}(z,Q^2)}{\pi \langle \bm{P}_{\perp,a \to h}^2 \rangle}e^{- \bm{P}_\perp^2/\langle \bm{P}_{\perp,a \to h}^2 \rangle}\ .
\label{e:fldep_gauss}
\end{align} 
%
Applying this hypothesis to the unpolarized structure function we obtain:
\begin{equation} 
F_{UU ,T} =  \sum_a e_a^2\ f_1^a(x,Q^2) D_1^{a \to h}(z,Q^2)\ \bigg[ \frac{e^{-\bm{k}_\perp^2/\langle \bm{k}_{\perp,a}^2 \rangle}}{\pi \langle \bm{k}_{\perp,a}^2 \rangle} \otimes \frac{e^{- \bm{P}_\perp^2/\langle \bm{P}_{\perp,a \rightarrow h}^2 \rangle}}{\pi \langle \bm{P}_{\perp,a \rightarrow h}^2 \rangle} \bigg] \ .
\label{e:F_UUT_fldep}
\end{equation}
Each convolution in Eq. \ref{e:F_UUT_fldep} results in a Gaussian function in $\bm{P}_{{hT}}$
\begin{equation}
\bigg[ \frac{1}{\pi \langle \bm{k}_{\perp,a}^2 \rangle}e^{-\bm{k}_\perp^2/\langle \bm{k}_{\perp,a}^2 \rangle} \otimes \frac{1}{\pi \langle \bm{P}_{\perp,a \rightarrow h}^2 \rangle}e^{- \bm{P}_\perp^2/\langle \bm{P}_{\perp,a \rightarrow h}^2 \rangle} \bigg] = \frac{x}{\pi \langle \bm{P}_{hT,a}^2 \rangle}e^{-\bm{P}_{hT}^2/\langle \bm{P}_{hT,a}^2 \rangle} \ ,
\label{e:Gauss_con}
\end{equation}
where the relation between the three variances is:
\begin{equation}
\langle \bm{P}_{hT,a}^2 \rangle = z^2 \langle \bm{k}_{\perp,a}^2 \rangle + \langle \bm{P}_{\perp,a \rightarrow h}^2 \rangle \ .
\label{e:fldep_transvmom_rel}
\end{equation}
This relation, peculiar of the Gaussian hypothesis, connects the measurable variable $\langle \bm{P}_{hT,a}^2 \rangle$ to the average square values of the intrinsic transverse momenta $\bm{k}_\perp$ and $\bm{P}_\perp$, not directly accessible by experiments.
%
Since $F_{UU,T}$ is a summation of Gaussians in $\bm{P}_{{hT}}$, it is not a Gaussian function any more. This is a crucial remark, since data point towards non-Gaussian functional forms for multiplicities.\\
Evaluating the cross sections using the flavor-dependent Gaussian hypothesis we obtain the following expression:
\begin{equation}
\begin{split}
m_{H}^h (x,z,Q^2,\bm{P}_{hT}^2) &= \frac{ \pi }{ \sum_{a} e_a^2 f_{1}^{H,a} (x,Q^2) }  \\
& \times \sum_{a} \biggl[ e_a^2 f_{1}^{H,a} (x,Q^2) D_{1}^{a \to h} (z,Q^2)\ \frac{ e^{ - \frac{ \bm{P}_{hT}^2 }{ z^2 \langle \bm{k}_{\perp,a}^2 \rangle + \langle \bm{P}_{\perp,a \to h}^2 \rangle }} }{ \pi (z^2 \langle \bm{k}_{\perp,a}^2 \rangle + \langle \bm{P}_{\perp,a \to h}^2 \rangle) } \biggr]  \ ,
\label{e:FDmult}
\end{split}  
\end{equation} 
where the Gaussian functions in $\bm{P}_{{hT}}$ result from the convolution in Eq. \ref{e:Gauss_con}.

\subsection{Assumptions concerning average transverse momenta}

In order to simplify the expression of the hadron multiplicity $m_H^h$ (Eq. \ref{e:FDmult}) and its phenomenological analysis, we collect all the flavors in three categories (up-valence quarks, down-valence quarks and flavor-symmetric sea quarks) and we impose isospin and charge conjugation symmetry on the transverse-momentum-dependent part of the fragmentation functions. 

Fragmentation processes in which the fragmenting parton is in the valence content of the detected hadron are defined to be {\em favored}. Otherwise the process is classified as {\em unfavored}. According to these arguments we are left with only two independent TMD parts in fragmentation functions: the favored one, describing 
\begin{equation}
u \to \pi^+ \equiv d \to \pi^- \equiv \bar{d} \to \pi^+ \equiv \bar{u} \to \pi^- \equiv \text{fav}\ ,    
\label{e:Pions.favored} 
\end{equation}
and the unfavored one, for the remaining processes
\begin{equation}
d \to \pi^+ \equiv u \to \pi^- \equiv \bar{u} \to \pi^+ \equiv \bar{d} \to \pi^- \equiv sea \to \pi^+ \equiv sea \to \pi^- \equiv \text{unf}\ .    
\label{e:Pions.unfavored} 
\end{equation} 
%
Our choice is to parametrize the variances with a flavor-dependent multiplicative factor and a flavor-independent kinematic dependence:
\begin{align} 
\langle \bm{k}_{\perp,i}^2 \rangle (x) = 
\langle \hat{\bm{k}}_{\perp,i}^2 \rangle \;  
\frac{(1-x)^{\alpha}\ x^{\sigma} }{ (1-\hat{x})^{\alpha}\ \hat{x}^{\sigma} } \, ,
\label{e:kT2_kin}
&&
\text{where }
\langle \hat{\bm{k}}_{\perp,i}^2 \rangle\equiv  \langle \bm{k}_{\perp,i}^2 \rangle
(\hat{x}),
\text{ and }
\hat{x}=0.1,
\end{align} 
where $i$ discriminates among up-valence, down-valence and sea quarks.
$\langle \hat{\bm{k}}_{T,i}^2 \rangle$, $\alpha$, $\sigma$, are free
parameters.
The parametrization of the width in the fragmentation is:
\begin{align}  
\langle \bm{P}_{\perp,j}^2 \rangle (z) &= \langle
\hat{\bm{P}}_{\perp,j}^2 \rangle 
\frac{ (z^{\beta} + \delta)\ (1-z)^{\gamma} }{ (\hat{z}^{\beta} + \delta)\
  (1-\hat{z})^{\gamma} } \, 
&&
\text{where }
\langle \hat{\bm{P}}_{\perp,j }^2 \rangle \equiv \langle \bm{P}_{\perp,j }^2 \rangle
(\hat{z}),
\text{ and }
\hat{z}=0.5.
\label{e:PT2_kin}
\end{align}  
where $\langle \hat{\bm{P}}_{T,j }^2 \rangle$, $\beta$, $\gamma$, and $\delta$ are free parameters and $j$ distinguish between favored and unfavored fragmentation processes.

\section{Data selection}
We choose the HERMES\cite{Airapetian:2012ki} data set without vector meson contributions, considering proton and deuteron targets and pions in the final state. 


The investigated kinematic region is $1.4\ \text{GeV}^2 < Q^2 < 9.2\ \text{GeV}^2$, $0.1 < z < 0.8$, $0.15 < | \bm{P}_{hT}  | < \sqrt{Q^2 / 3}\ \text{GeV}$ .
Because of the limited explored $Q^2$ region we can carry out the analysis at a fixed $Q^2 = 2.4\ \text{GeV}^2$ in Eq. \ref{e:FDmult}, neglecting evolution effects.
A more detailed description of the cuts is available in Ref. \refcite{Signori:2013mda}.


\section{Fitting procedure}

The approach consists in creating ${\cal M}$ replicas of HERMES data set. In each replica (denoted by the index $r$), each data point is shifted by a Gaussian noise with the same variance as the measurement. Each replica, therefore, represents a possible outcome of an independent experimental measurement, which we denote by $m_{H, r}^{h}$. 

For each replica we minimize the following error function:
\begin{equation}
E_r^2(\{p\})=\sum_{i, H, h} 
\frac{\Bigl(m_{H,i,r}^{h} - m_{H,i,  \mbox{\tiny theo}}^{h}(\{p\})\Bigr)^2}
        {\Bigl(\Delta m_{H,i, \mbox{\tiny stat}}^{h} \Bigr)^2+\Bigl(\Delta m_{H,i, \mbox{\tiny sys}}^{h} \Bigr)^2+\Bigl(\Delta m_{H,i, \mbox{\tiny theo}}^{h} \Bigr)^2}  \ . 
\label{e:MC_chi2}
\end{equation}
The sum runs over the experimental points (indicated by kinematic bin $i$, target $H$, and final-state hadron $h$); $\{p\}$ denotes the vector of parameters. The
theoretical contribution to the error on the multiplicities mainly comes from the propagation to Eq. \ref{e:FDmult} of the error on the collinear fragmentation functions $D_1(z)$\cite{Epele:2012vg}.

From the minimization of all replicas we can calculate ${\cal M}$ different vectors of best values for the fit parameters ($\{ p_{0r}\},\; r=1,\ldots {\cal M}$). 
The agreement of the ${\cal M}$ theoretical outcomes with the original data is better expressed in terms of a $\chi^2$ function calculated not with respect to the replica, but with respect to the original data set.

\section{Phenomenological results from HERMES}
\label{s:results}
In the next paragraphs we describe the results available from the flavor-dependent and independent fits of ${\cal M}=200$ sets of multiplicities built from Hermes data.
In both the investigations the fit function is given in Eq. \ref{e:FDmult}, whereas the number of its parameters changes in the two analysis (10 in the flavor-dependent case, 7 in the flavor-independent one).

\subsection{Flavor-dependent fit}

\begin{figure}
\centering
\includegraphics[width=7cm]{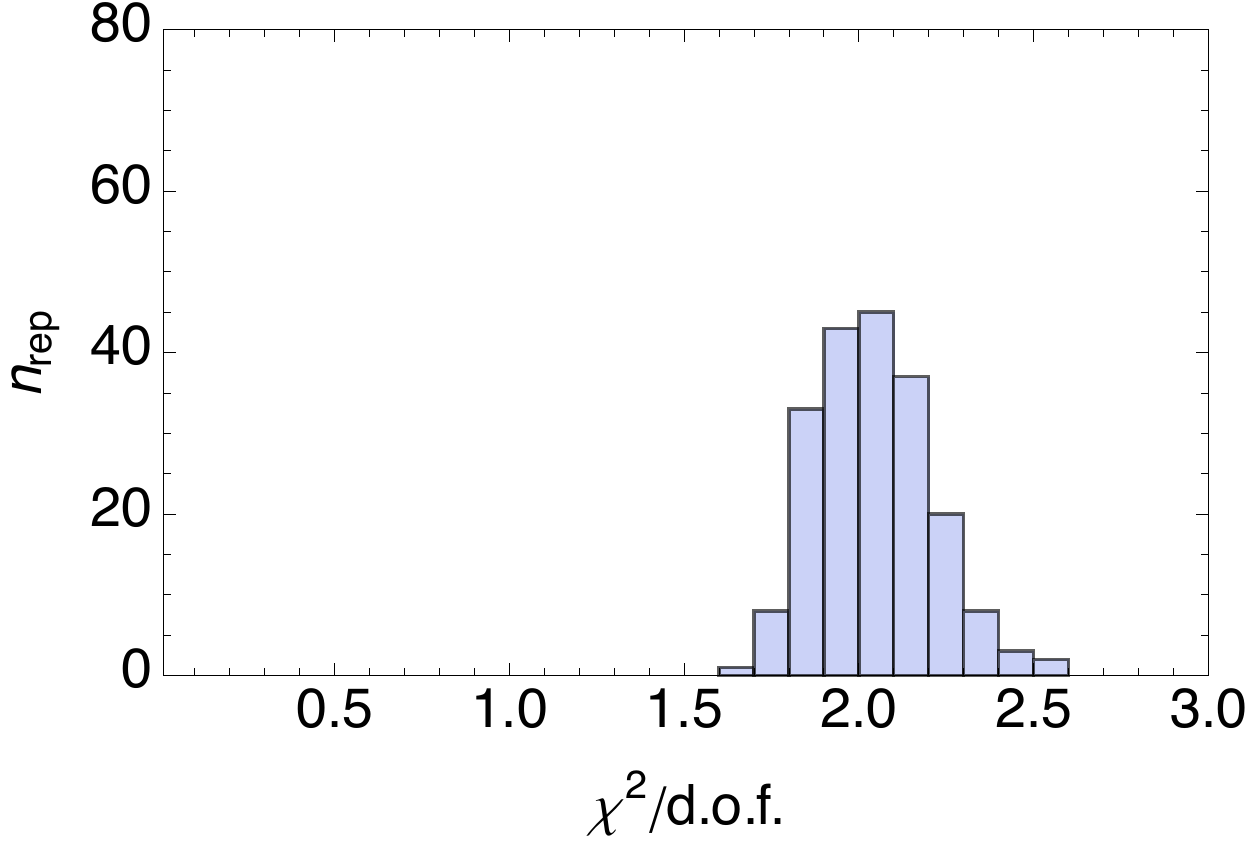}
\caption{$\chi^2$/d.o.f. for the flavor-dependent fit.}
\label{f:fd_chi20dof}
\end{figure}
\begin{figure}
\centering
\begin{tabular}{ccc}
\includegraphics[width=6.3cm]{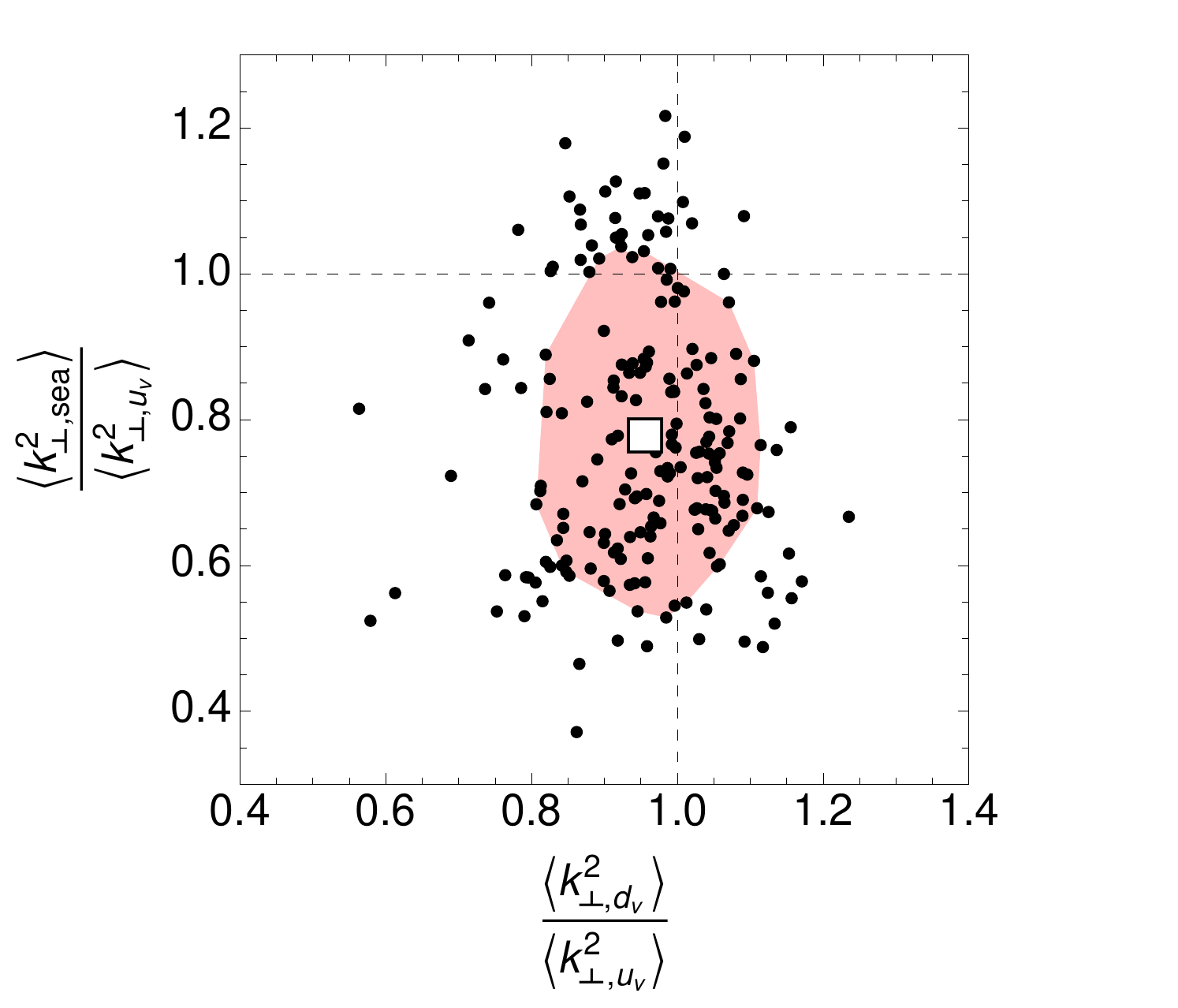}
&\hspace{0.01cm}
&
\includegraphics[width=6.3cm]{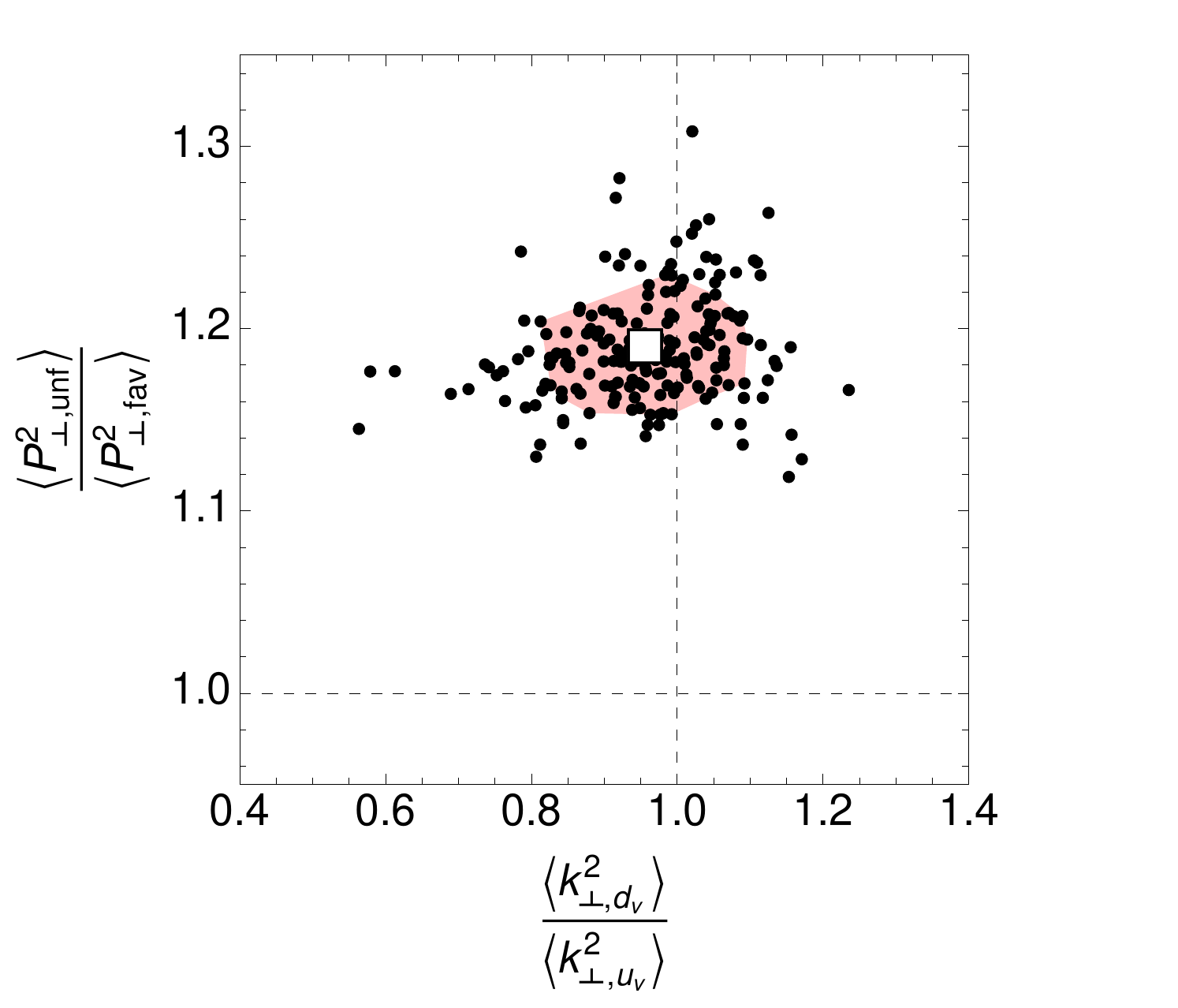}
\\
(a) && (b)
\end{tabular}
\caption{(a) Distribution of the values of the ratios 
$\langle \bm{k}^2_{\perp, d_v} \rangle / \langle \bm{k}^2_{\perp, u_v}\rangle$ vs. 
$\langle \bm{k}^2_{\perp, \text{sea}}\rangle / \langle \bm{k}^2_{\perp, u_v}\rangle$ obtained
from fitting 200 replicas of the original data points. The white squared box indicates the center of the 68\% confidence
interval for each ratio.
The shaded area represents the two-dimensional 68\% confidence
region around the white box. 
The dashed lines correspond to the ratios being unity; their crossing
point corresponds to the result with no flavor dependence. For most of the points, $\langle \bm{k}^2_{\perp,\text{sea}} \rangle < \langle \bm{k}^2_{\perp, d_v}\rangle \lesssim \langle \bm{k}^2_{\perp, u_v}\rangle$. 
(b) Same as previous panel, with distribution of the values of the ratios 
$\langle \bm{P}^2_{\perp, \text{unf}}\rangle / \langle \bm{P}^2_{\perp, \text{fav}}\rangle$ vs. 
$\langle \bm{k}^2_{\perp, d_v}\rangle / \langle \bm{k}^2_{\perp, {u_v}}\rangle$. For all points $\langle \bm{P}^2_{\perp, \text{fav}}\rangle < \langle \bm{P}^2_{\perp, \text{unf}}\rangle$.
}
\label{f:DoverU_SoverU_pions_only}
\end{figure}

The $\chi^2$/d.o.f. of the ${\cal M}$ replicas are distributed as in Fig. \ref{f:fd_chi20dof}. Their mean value and standard deviation are:
\begin{equation}
\chi^2/\text{d.o.f.} = 2.04 \pm 0.16
\label{e:chi20dof_FD}
\end{equation}
These numbers may not look satisfactory. But it must be reminded that the collected HERMES bins contain a very large statistics and, moreover, that large $\chi^2$ are obtained in the fit of the corresponding collinear multiplicities (for detail see Ref. \refcite{Signori:2013mda}).
%
Mean values and standard deviations of best values for $\langle \hat{\bm{k}}_{\perp,i}^2 \rangle$ (Eq. \ref{e:kT2_kin}) are:
\begin{align}
&
\langle \hat{\bm{k}}_{\perp,\text{u}_{\text{v}}}^2 \rangle = 0.34 \pm 0.10 \ \text{GeV}^2 \\
&
\langle \hat{\bm{k}}_{\perp,\text{d}_{\text{v}}}^2 \rangle = 0.33 \pm 0.12 \ \text{GeV}^2 \\
&
\langle \hat{\bm{k}}_{\perp,\text{sea}}^2 \rangle = 0.27 \pm 0.12 \ \text{GeV}^2 .
\label{e:best_NkT}
\end{align}
Their distribution for all the replicas is described in Fig. \ref{f:DoverU_SoverU_pions_only}. The $\alpha$ and $\sigma$ parameters were fixed in each replica extracting a uniform random number in $(0,2)$ and $(-0.3,0.1)$, respectively; accordingly, their mean value and standard deviation are:
\begin{align}
&
\alpha = 0.92 \pm 0.57 \ ,
&
\sigma = -0.10 \pm 0.12 \ .
\label{e:best_alpha}
\end{align}
%
%
Mean values and standard deviations of $\langle \hat{\bm{P}}_{\perp,j}^2 \rangle$ (Eq. \ref{e:PT2_kin}) are:
\begin{align}
&
\langle \hat{\bm{P}}_{\perp,\text{fav}}^2 \rangle = 0.17 \pm 0.03 \ \text{GeV}^2 \\
&
\langle \hat{\bm{P}}_{\perp,\text{unf}}^2 \rangle = 0.20 \pm 0.03 \ \text{GeV}^2 \ .   
\label{e:best_NPT}
\end{align}
Fig. \ref{f:DoverU_SoverU_pions_only} shows their distributions for all the replicas.
$\beta$, $\gamma$ and $\delta$ are free parameters; their mean values and standard deviations are:
\begin{align}
&
\beta = 1.54 \pm 0.27 \ \ \ \  \gamma = 1.11 \pm 0.60 \ \ \ \  \delta = 0.15 \pm 0.05 \ .
\label{e:best_alpha}
\end{align}
%
%
\begin{figure}
\centering
\includegraphics[width=13.5cm]{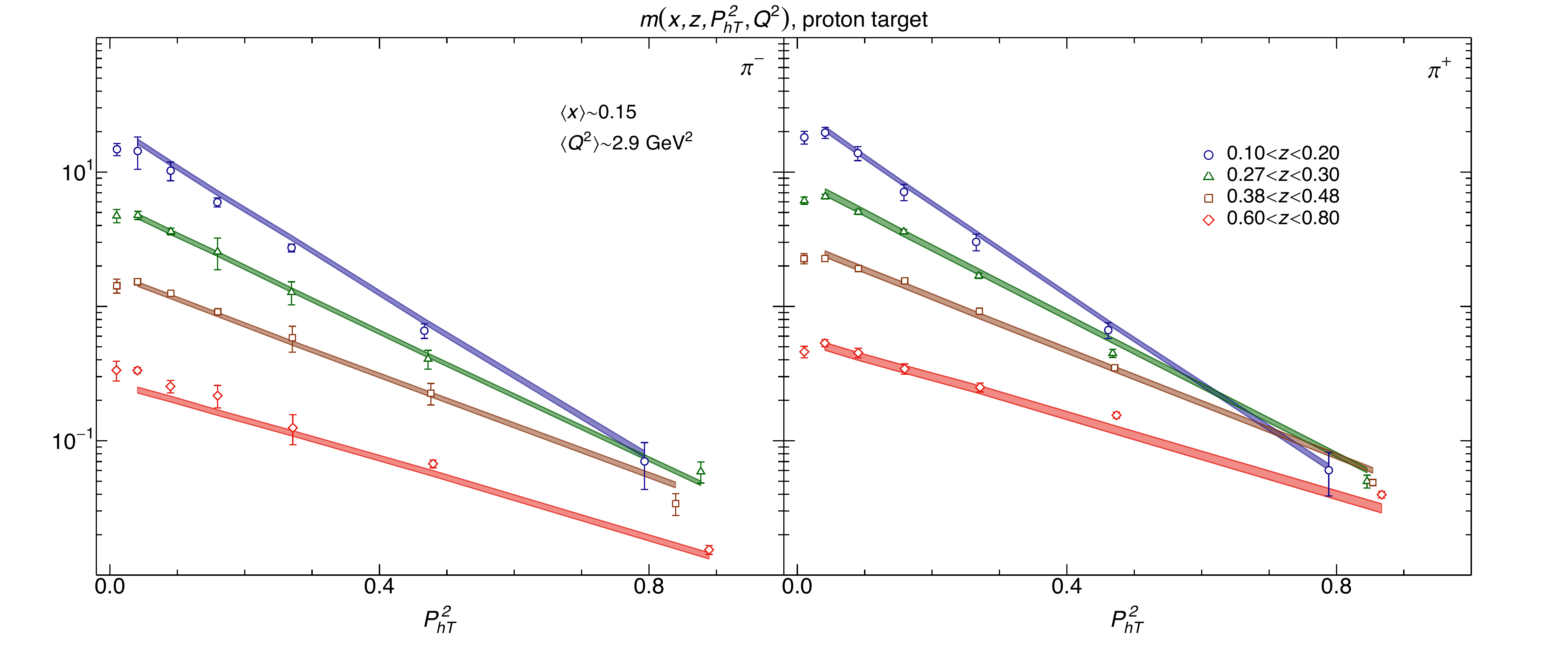}
\caption{The global agreement between Hermes data (with proton target) at $\langle x \rangle = 0.15$, $\langle Q^2 \rangle = 2.9\ \text{GeV}^2$ and the flavor-dependent Gaussian model. The width of each theoretical band corresponds to a $68\%$ confidence level.}
\label{f:fd_Ppions}
\end{figure}
\begin{figure}
\centering
\includegraphics[width=13.5cm]{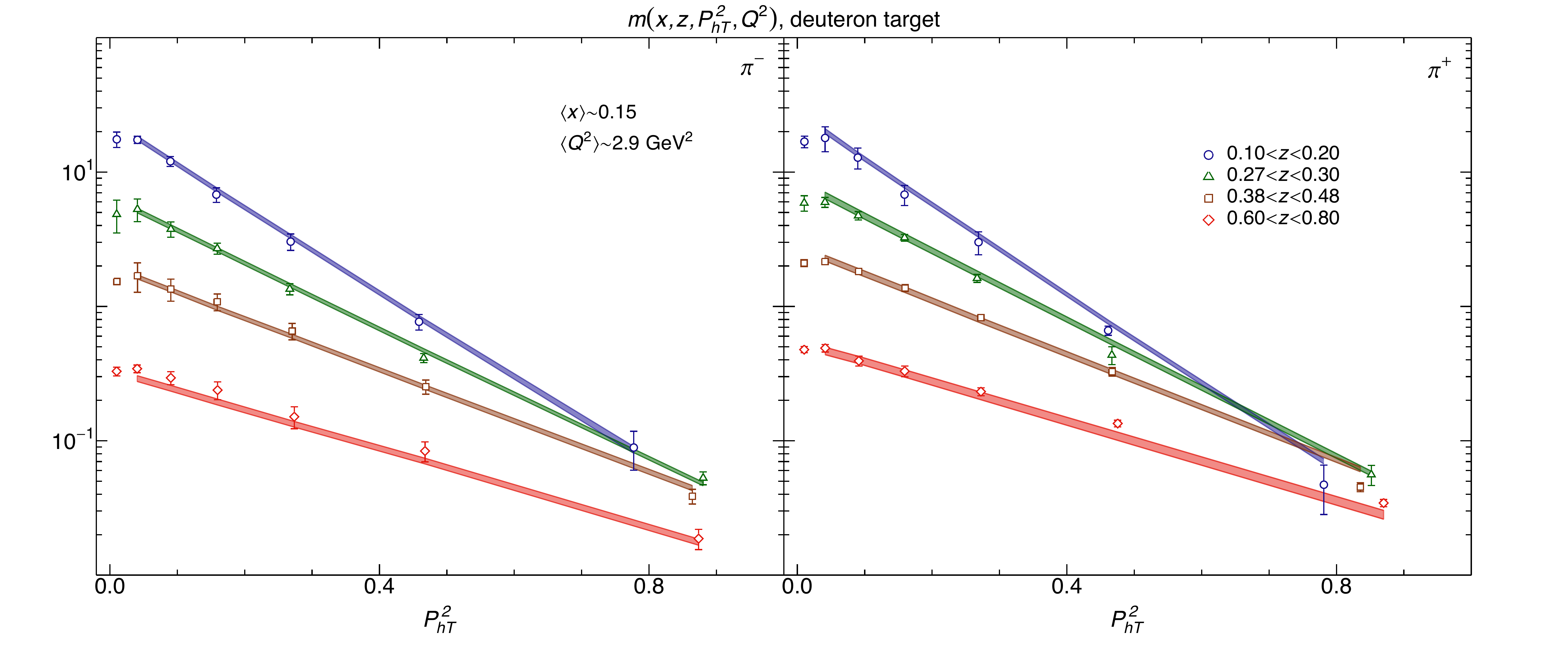}
\caption{Same as in Fig. \ref{f:fd_Ppions} but for Hermes data with a deuteron target.}
\label{f:fd_Dpions}
\end{figure}
%
All the $\langle \hat{\bm{k}}_{\perp,i}^2 \rangle$ parameters involved in the analysis are correlated, and the same holds for $\langle \hat{\bm{P}}_{\perp,j}^2 \rangle$. The widths in distribution and fragmentation functions, instead, are anti-correlated, as suggested by Eq. \ref{e:fldep_transvmom_rel}.
%
%
%
%
The agreement between HERMES data and the flavor-dependent model is nicely shown by Fig. \ref{f:fd_Ppions} and Fig. \ref{f:fd_Dpions}.
From this analysis, it emerges that valence quarks are on average similar, and tend to have a wider distribution than the sea quarks. However, for many replicas the ratio deviates from the mean value by more than 20\%. At variance, the unfavored fragmentation has clearly a larger mean square transverse momentum than the favored one.
The flavor-dependent model fits data better than the flavor-independent one, altough its $\chi^2$/d.o.f. is not strikingly smaller.

\subsection{Flavor-independent fit}
The agreement achieved with the flavor-independent model is slightly worse than the previous one. Mean value and standard deviation of the $\chi^2$/d.o.f. are:
\begin{equation}
\chi^2/\text{d.o.f.} = 2.15 \pm 0.16 \ .
\label{e:chi20dof_FI}
\end{equation}
%
In this analysis we have only one Gaussian function (with variance $\langle \bm{k}_\perp^2 \rangle$) for the distribution and one Gaussian function (with variance $\langle \bm{P}_\perp^2 \rangle$) for the fragmentation function. Here we summarize mean values and standard deviations of fit parameters:
\begin{align}
&
\langle \hat{\bm{k}}_\perp^2 \rangle = 0.22 \pm 0.10 \ \text{GeV}^2 \ \ \ \ \
\langle \hat{\bm{P}}_\perp^2 \rangle = 0.20 \pm 0.03 \ \text{GeV}^2
\label{e:best_mult_FI}
\end{align}
\begin{align}
&
\ \ \ \ \ \ \ \ \ \ \ \alpha = 0.96 \pm 0.59 \ \ \ \  \sigma = -0.11 \pm 0.12
\nonumber \\ &
\beta = 1.42 \pm 0.24 \ \ \ \  \gamma = 0.74 \pm 0.42 \ \ \ \  \delta = 0.19 \pm 0.04 \ .
\label{e:best_kin_FI}
\end{align}

\section{Conclusions}
\label{s:conclusions}

Results indicate that the model with flavor-dependent Gaussian TMDs fits data slightly better than the same model without flavor dependence, hence the latter cannot be a priori excluded.
However we clearly observe wider Gaussians for unfavored fragmentation functions compared to the favored ones.
We also observe a bigger mean average square transverse momentum for valence quark distributions compared to sea quarks, and similar up and down valence distributions. 
However, in this case the values of transverse momenta are so spread that finding a sea quark distribution larger than the valence one is statistically relevant, as well as finding valence down and up quarks that differ by 20\%. 
With this analysis we provide for the first time the phenomenological tools to explore the flavor dependence of the transverse motion of partons inside hadrons.
Moreover, we also account for the possible non-Gaussian behavior of the unpolarized structure function $F_{UU,T}$.
A more extended and complete analysis has been described in Ref. \refcite{Signori:2013mda} considering also kaons in the final state.

\section*{Acknowledgments}
This research is part of the program of the ÒStichting voor Fundamenteel Onderzoek der Materie (FOM)Ó, which is financially supported by the ÒNederlandse Organisatie voor Wetenschappelijk Onderzoek (NWO)Ó.
It is also partially supported by the European Community through the
Research Infrastructure Integrating Activity ``HadronPhysics3'' (Grant Agreement
n. 283286) under the 7th Framework Programme.
Discussions with Maarten Buffing, Marco Contalbrigo, Marco Guagnelli, the HERMES collaboration, Piet Mulders, Barbara Pasquini, Gunar Schnell, Marco Stratmann are gratefully acknowledged.

\bibliographystyle{utphys}
\bibliography{mybiblio}

\end{document}